\documentstyle[psfig]{article}
\newcommand{\eq}{\begin{equation}}
\newcommand{\eqe}{\end{equation}}

\newcommand{\pa}{\partial}
\newcommand{\vep}{\varepsilon}
\newcommand{\vp}{\varphi}
\newcommand{\phit}{\phi^4_4}

\newcommand{\De}{\Delta}
\newcommand{\La}{\Lambda}

\newcommand{\ga}{\gamma}
\newcommand{\be}{\beta}
\newcommand{\Lao}{\Lambda_0}
\newcommand{\de}{\delta}

\newcommand{\Lll}{L^{\Lambda,\Lambda_0}}
\newcommand{\Llol}{L^{\Lambda_0,\Lambda_0}}
\newcommand{\cLlln}{{\cal L}^{\Lambda,\Lambda_0}_{l,n}}
\newcommand{\ccL}{{\cal L}}
\newcommand{\cLl}{{\cal L}^{\Lambda}}
\newcommand{\cLll}{{\cal L}^{\Lambda,\Lambda_0}}

\newcommand{\cLol}{{\cal L}^{0,\Lambda_0}}

\newcommand{\cLlol}{{\cal L}^{\Lambda_0,\Lambda_0}}

\begin{document}

\message{reelletc.tex (Version 1.0): Befehle zur Darstellung |R  |N, Aufruf=
z.B. \string\bbbr}
%
%
\message{reelletc.tex (Version 1.0): Befehle zur Darstellung |R  |N, Aufruf=
z.B. \string\bbbr}
\font \smallescriptscriptfont = cmr5
\font \smallescriptfont       = cmr5 at 7pt
\font \smalletextfont         = cmr5 at 10pt
\font \tensans                = cmss10
\font \fivesans               = cmss10 at 5pt
\font \sixsans                = cmss10 at 6pt
\font \sevensans              = cmss10 at 7pt
\font \ninesans               = cmss10 at 9pt
\newfam\sansfam
\textfont\sansfam=\tensans\scriptfont\sansfam=\sevensans
\scriptscriptfont\sansfam=\fivesans
\def\sans{\fam\sansfam\tensans}
\def\bbbr{{\rm I\!R}} 
\def\bbbn{{\rm I\!N}} 
\def\bbbE{{\rm I\!E}} 
\def\bbbm{{\rm I\!M}}
\def\bbbh{{\rm I\!H}}
\def\bbbk{{\rm I\!K}}
\def\bbbd{{\rm I\!D}}
\def\bbbp{{\rm I\!P}}
\def\bbbone{{\mathchoice {\rm 1\mskip-4mu l} {\rm 1\mskip-4mu l}
{\rm 1\mskip-4.5mu l} {\rm 1\mskip-5mu l}}}
\def\bbbc{{\mathchoice {\setbox0=\hbox{$\displaystyle\rm C$}\hbox{\hbox
to0pt{\kern0.4\wd0\vrule height0.9\ht0\hss}\box0}}
{\setbox0=\hbox{$\textstyle\rm C$}\hbox{\hbox
to0pt{\kern0.4\wd0\vrule height0.9\ht0\hss}\box0}}
{\setbox0=\hbox{$\scriptstyle\rm C$}\hbox{\hbox
to0pt{\kern0.4\wd0\vrule height0.9\ht0\hss}\box0}}
{\setbox0=\hbox{$\scriptscriptstyle\rm C$}\hbox{\hbox
to0pt{\kern0.4\wd0\vrule height0.9\ht0\hss}\box0}}}}

\def\bbbe{{\mathchoice {\setbox0=\hbox{\smalletextfont e}\hbox{\raise
0.1\ht0\hbox to0pt{\kern0.4\wd0\vrule width0.3pt height0.7\ht0\hss}\box0}}
{\setbox0=\hbox{\smalletextfont e}\hbox{\raise
0.1\ht0\hbox to0pt{\kern0.4\wd0\vrule width0.3pt height0.7\ht0\hss}\box0}}
{\setbox0=\hbox{\smallescriptfont e}\hbox{\raise
0.1\ht0\hbox to0pt{\kern0.5\wd0\vrule width0.2pt height0.7\ht0\hss}\box0}}
{\setbox0=\hbox{\smallescriptscriptfont e}\hbox{\raise
0.1\ht0\hbox to0pt{\kern0.4\wd0\vrule width0.2pt height0.7\ht0\hss}\box0}}}}

\def\bbbq{{\mathchoice {\setbox0=\hbox{$\displaystyle\rm Q$}\hbox{\raise
0.15\ht0\hbox to0pt{\kern0.4\wd0\vrule height0.8\ht0\hss}\box0}}
{\setbox0=\hbox{$\textstyle\rm Q$}\hbox{\raise
0.15\ht0\hbox to0pt{\kern0.4\wd0\vrule height0.8\ht0\hss}\box0}}
{\setbox0=\hbox{$\scriptstyle\rm Q$}\hbox{\raise
0.15\ht0\hbox to0pt{\kern0.4\wd0\vrule height0.7\ht0\hss}\box0}}
{\setbox0=\hbox{$\scriptscriptstyle\rm Q$}\hbox{\raise
0.15\ht0\hbox to0pt{\kern0.4\wd0\vrule height0.7\ht0\hss}\box0}}}}

\def\bbbt{{\mathchoice {\setbox0=\hbox{$\displaystyle\rm
T$}\hbox{\hbox to0pt{\kern0.3\wd0\vrule height0.9\ht0\hss}\box0}}
{\setbox0=\hbox{$\textstyle\rm T$}\hbox{\hbox
to0pt{\kern0.3\wd0\vrule height0.9\ht0\hss}\box0}}
{\setbox0=\hbox{$\scriptstyle\rm T$}\hbox{\hbox
to0pt{\kern0.3\wd0\vrule height0.9\ht0\hss}\box0}}
{\setbox0=\hbox{$\scriptscriptstyle\rm T$}\hbox{\hbox
to0pt{\kern0.3\wd0\vrule height0.9\ht0\hss}\box0}}}}

\def\bbbs{{\mathchoice
{\setbox0=\hbox{$\displaystyle     \rm S$}\hbox{\raise0.5\ht0\hbox
to0pt{\kern0.35\wd0\vrule height0.45\ht0\hss}\hbox
to0pt{\kern0.55\wd0\vrule height0.5\ht0\hss}\box0}}
{\setbox0=\hbox{$\textstyle        \rm S$}\hbox{\raise0.5\ht0\hbox
to0pt{\kern0.35\wd0\vrule height0.45\ht0\hss}\hbox
to0pt{\kern0.55\wd0\vrule height0.5\ht0\hss}\box0}}
{\setbox0=\hbox{$\scriptstyle      \rm S$}\hbox{\raise0.5\ht0\hbox
to0pt{\kern0.35\wd0\vrule height0.45\ht0\hss}\raise0.05\ht0\hbox
to0pt{\kern0.5\wd0\vrule height0.45\ht0\hss}\box0}}
{\setbox0=\hbox{$\scriptscriptstyle\rm S$}\hbox{\raise0.5\ht0\hbox
to0pt{\kern0.4\wd0\vrule height0.45\ht0\hss}\raise0.05\ht0\hbox
to0pt{\kern0.55\wd0\vrule height0.45\ht0\hss}\box0}}}}

\def\bbbz{{\mathchoice {\hbox{$\sans\textstyle Z\kern-0.4em Z$}}
{\hbox{$\sans\textstyle Z\kern-0.4em Z$}}
{\hbox{$\sans\scriptstyle Z\kern-0.3em Z$}}
{\hbox{$\sans\scriptscriptstyle Z\kern-0.2em Z$}}}}

\title{Irrelevant Interactions without Composite Operators -\\
 A Remark on the Universality of second order Phase Transitions} \maketitle

\centerline{Ch. Kopper and W. Pedra
\footnote{Stagiaire}}
\centerline{Centre de Physique Th{\'e}orique, CNRS UPR 14}
\centerline{Ecole Polytechnique} \centerline{91128 Palaiseau Cedex,
  FRANCE} \vskip 1cm \medskip
\noindent{\bf Abstract}

We study the critical behaviour of symmetric $\phi^4_4$ theory
including irrelevant
terms of the form $\phi^{4+2n}/\Lao^{2n}\,$ 
in the bare action,  where $\Lao\,$ is
the UV cutoff (corresponding e.g. to the inverse 
lattice spacing for a spin system).
The main technical tool is renormalization theory based on the flow
equations of the renormalization group which permits
to establish  the required convergence statements 
in generality and rigour. As a consequence the effect of irrelevant 
terms on the critical behaviour may be studied to any order
without using renormalization theory for composite operators.
This is a technical simplification and seems preferable from the physical
point of view. In this short note we restrict for simplicity
to the symmetry class of the Ising model, i.e.
one component $\phi^4_4$ theory. The method is general, however. 
\maketitle

\section{Introduction}
One of the great achievements of theoretical physics in the 70's
was the unification of concepts and ideas from quantum field theory 
and statistical mechanics through the Wilson renormalization group
[WiKo]. In particular renormalized perturbation theory
was applied  successfully 
to the study of second order phase transitions and to the
calculation of critical exponents [BGZ, Amit, ZJ]. 
One of the challenging conceptual problems was the question of
universality, i.e. to realize why large classes
of theories, specified essentially by the respective Hamiltonians
should give
rise to the same critical behaviour characterized through the
critical exponents. Experimentally those depend only 
on dimensionality and symmetry
but not on details of the dynamics. Modifications of the 
Hamiltonians  thus 
should lead only
to subleading corrections. We restrict our explicit presentation 
to one of the simplest and bestknown classes, that of the Ising model.
The method is general, however. 
Passing to the continuous description which should be viable for
correlation lengths $\xi$ much larger than the lattice spacing, i.e.
in the vicinity of the critical point, the symmetry class 
of the Ising model is presented by $\phi^4$ theory, symmetric
under $\phi \to -\phi$.   
The standard action at the scale of the UV cutoff $\Lao$, 
corresponding to the inverse lattice spacing in position space, is then
\eq
L^{0}\,=\,
\int \left(a\, \phi^2(x)\,+\,b
(\partial_{\mu}\phi)^2(x)\,+\,c\,\phi^4(x)\,\right)\;d^4x\;
\,.
\label{phi4}
\eqe
If we restrict ourselves to perturbation theory the constants 
$a,\ b,\ c\,$ are to be viewed as power series in the renormalized
coupling $g\,$ or in $\hbar\,$. 
In the standard notation this expression is rewritten as
\eq
L'^{0}\,=\,
\int \left({Z\over 2} (\partial_{\mu}\phi)^2(x)\,+\,
{Z\over 2} (m^2+\de m^2) \phi^2(x)\,+\,
{g_0 \over 4~!} Z^2\,\phi^4(x)\, \right)\,d^4x\; ,
\label{phi41}
\eqe
where $L'^{0}$ also includes the term of order 0 in perturbation
theory, that is to say
\eq
L'^{0}\,=\, L^{0}\,+\,\int \left({1\over 2}(\partial_{\mu}\phi)^2(x)\,+\,
{1\over 2} m^2 \phi^2(x)\right)\,d^4x\ .
\eqe 
The field $\phi\,$ corresponds to the renormalized field. 
In (\ref{phi41}) we
introduced the standard notation for the wave function renormalization
$Z\,$ and the mass counterterm $\de m\,$ as well as
the bare coupling $g_0\,$.
We restrict to  the four dimensional theory which also serves to study
lower dimensional theories through the 
$\vep\,$-expansion.\footnote{Dimensional regularization cannot be naturally
 accommodated   in the flow equation framework. Still the associated minimal 
renormalization schemes should be implementable. This has been shown 
for analytic regularization [KoSm].}
Starting from the Ising model Hamiltonian on a cubic lattice one 
arrives at the 
action (\ref{phi4}) on performing block spin transformations,
expanding in local terms  and
passing to the continuous limit, on neglecting all irrelevant terms,
i.e. those of mass dimension larger than 4. The aim of this paper is
to show that this is justified indeed when analysing long distance
phenomena near the critical point. This means that the dominant
contributions to the correlation functions near the critical 
point are obtained from (\ref{phi4}) for suitable choices 
of $  a,\ b,\ c\,$. 
More precisely we will add a finite sum
\eq
A(\phi)\,=\,
\int \sum_{n=1}^N {Z^{2+n}\over \Lao ^{2n}}\,
  {g_{4+2n} \over \ (4+2n)! } \,
\phi ^{4+2n}(x)\: d^4x 
\label{irre}
\eqe
to (\ref{phi4}). Here the UV cutoff $\Lao$ appears naturally when 
expanding in
local terms, by dimensional analysis. This
means that the couplings $g_{4+2n}$ are dimensionless 
(in $d=4\,$).\footnote{We choose conventions such that a factor of $Z^{2+n}$
appears in front of $g_{4+2n}\,$, which will somewhat simplify the
notation later. Note that $Z\,$ will depend on the couplings
$g_{4+2n}\,$. In particular it will also be
different from $1$ for  $g=0\,$, if some of the $g_{4+2n}\,$ do not vanish.} 
Since the statements of renormalization theory are generally of perturbative
nature, i.e. valid on  formal expansion in the couplings $g$, they 
require small values of those to be reliable. When including
(\ref{irre}) the question arises how the size of the irrelevant 
couplings compares to that of the original $\phi ^4\,$ coupling $g\,$.
Here of course different situations may arise and can be analysed. 
Later on we will
regard the situation where they are chosen
such that the loop expansion remains valid, which means generally that
\eq
g_{4+2n} \sim \,g^{n+1}\,.
\label{boucle}
\eqe
The expansion with respect to local terms 
also produces higher dimensional terms of the form
$(\phi ^n \pa ^w \phi ^m)(x)\,$, which contain $|w|\,$ derivatives with
respect to  the coordinates $x\,$. 
Starting from a cubic lattice only terms respecting rotational
symmetry, i.e. invariant under the Euclidean group, should appear
in the continuum limit, i.e. when approaching the critical region.
Furthermore in (\ref{irre}) only terms invariant
under $\phi \to -\phi\,$ are generated
if $\bbbz_2$-symmetry is unbroken. For shortness of notation 
we restrict to (\ref{irre}), inclusion
of derivative terms would only lead to minor changes.  
In the explicit treatment we will even limit ourselves to a
single  insertion $\sim g_6  \phi ^6\,$, for simplicity of notation. 
 
The effects of irrelevant terms have of course been studied
extensively in the literature [Weg, BGZ] 
and can be found in textbooks, e.g. [ZJ, Ch.26]. 
In the field theory approach these terms 
were analysed by  renormalization theory 
for composite operators, as it existed in the early seventies [Zim].
Treating e.g. $\phi^6\,$ as a composite operator insertion
means that one restricts to Green functions 
carrying at most one insertion of this $\phi^6\,$ 
term.\footnote{ It is possible to go beyond one insertion.
But then the  number of renormalization conditions one has to fix
increases with the number of insertions, corresponding to the fact
that a $\phi^6\,$ theory is nonrenormalizable.}
Then one has to fix renormalization conditions 
for the inserted Green functions
up to dimension six (thus on the two, four and six
point functions and on derivatives of the two and four point functions). 
The general and probably optimal bound on the coefficient of the 
term $\phi ^6$ in $L^0$ is then of the form ${\cal P }(\log\Lao)\,$, i.e.
a polynomial in logarithms of $\Lao$  
-as has been shown e.g. in [KeKo1]- and not 
$\sim \Lao ^{-2}\,{\cal P}(\log\Lao)\,$ as in (\ref{irre}). 
Otherwise stated this means that
in general it is not possible to find out the renormalization conditions
which would give a bound $\sim \Lao ^{-2}$, since the associated
dynamical system is unstable. 
From the physical point of view it seems therefore 
perferable to start directly from
the modification of the {\it bare}   action as in (\ref{irre}), and 
to perform the renormalization for this theory. 
We note however that it is not really possible to study the question
in such a way that the {\it only} change in the bare action consists in
adding the term  $\sim \phi^6\,$ to it. The counterterms $\sim \phi^4\,$ 
and $\sim \phi^2\,$ change at the same time if we keep the
renormalization conditions fixed. This phenomenon corresponds to what
is called operator mixing in the theory of composite operators. 
However, whereas these renormalization conditions generally 
are  related to the 
physical parametrization of the theory near the critical region
and thus accessible to experiment, this seems  not to be the case 
for the Green functions carrying  e.g. $ \phi^6\,$-insertions.
Our results are such
that we may study an arbitrary number of irrelevant insertions 
for a fixed set of renormalization conditions.
Thus the study of these insertions is
generalized and simplified at the same time as compared to
composite operator theory. 

Renormalizability proofs based on the renormalization group 
are conceptually simple and rigorous
and give a transparent 
view on the universality of critical behaviour, in showing that the 
modification of the action by irrelevant terms does not influence
(the dominant part of) the critical behaviour. 
In this note we would like to make
this explicit for the simplest case, the universality class of the
Ising model. In the next section we will present the 
required results on the renormalizability of the theories with and 
without $\phi ^6\,$-insertion, in particular in the critical region.  
In the last section we also use the (standard) renormalization group
equations to analyse the subdominant contributions of the irrelevant
insertion to the critical behaviour.

\section{Renormalization of $\phi^4_4$ theory with irrelevant terms}

Renormalization theory as we are going to use it here is based on the
flow equations of the renormalization group due to Wilson [WiKo],
and particularly  to Polchinski [Pol] as regards the application to
the perturbative renormalization problem. The flow equation is obtained
by successively integrating out momenta in the (regularized) theory
starting from the UV cutoff $\Lao$ down to the scale $\La < \Lao$. 
 The final renormalized theory is obtained on taking the limits 
$\Lao \to \infty\,$ and $\La \to 0\,$. Its differential form 
can be obtained when deriving with respect to $\La\,$ 
the generating functional of the
connected (free propagator) amputated Green functions (CAG) 
of the theory with
momenta restricted to lie between $\La\,$ and $\Lao\,$. 
The scales   
$\La\,$ and $\Lao\,$ enter through the regularized propagator,
which for the massive theory takes the form
\eq
C^{\La,\Lao}(p)\,=\, {1 \over  p^2+m^2} 
\{ e^{- {p^2+m^2 \over \Lao^2}} -e^{- {p^2+m^2 \over \La^2}} \}\ .
\label{prop}
\eqe
Its  Fourier transform is  
\eq
\hat{C}^{\La,\Lao}(x)\,=\, 
\int_p C^{\La,\Lao}(p) \,e^{ipx}\ .
\label{propx}
\eqe
\[
\mbox{ We use the conventions: }\quad 
\int_p~:= \int_{\bbbr^4}  {d^4 p \over (2\pi)^4}\,,  \quad 
\phi(x) = \int_p  \vp(p) \,e^{ipx}\,,\quad
{\de \over \de\phi(x)} =
(2\pi)^4 
\int_p {\de \over \de \vp(p)}\, e^{-ipx}\ .
\footnote{For our 
purposes the "fields" $\phi(x)$ 
may be assumed to live in
the Schwartz space ${\cal S}(\bbbr^4)$.}
\]
For finite $\Lao$ and in finite volume the theory can be given
rigorous meaning starting from the functional integral
\eq
e^{- {1\over \hbar} (L^{\La,\Lao}(\phi)+ I^{\La,\Lao})}
\,=\, 
\int \, d\mu_{\La,\Lao}(\Phi) \; 
e^{- {1\over\hbar} \Llol(\phi\,+\,\Phi)} \ ,
\label{funcin}
\eqe
where the factors of $\hbar\,$ have been introduced to allow for a consistent
loop expansion in the sequel which permits us to stay
with a single expansion parameter in the presence of two coupling constants.
In (\ref{funcin}) $\,d\mu_{\La,\Lao}(\Phi) $ denotes the (translation
invariant) Gaussian measure with covariance $\hbar \hat{C}^{\La,\Lao}(x)\,$.
The normalization factor $\,e^{-{1\over\hbar} I^{\La,\Lao}}\, $ is
due to vacuum contributions. It
diverges in infinite volume so that we can take the infinite volume
limit only when it has been eliminated. We do not make the finite
volume explicit here since it plays no role in the sequel.  
One may convince oneself that $\Lll(\phi) \,$ is equal to
\eq
\Lll(\phi) \,=\,
-\ln Z^{\La,\Lao}((\hat{C}^{\La,\Lao})^{-1}\,\phi)\,+\, 
1/2\,\langle \phi,\,(\hat{C}^{\La,\Lao})^{-1}\phi  \rangle
\ . 
\eqe
Here $\,Z^{\La,\Lao}(j)\,$ is the (standard notation for the) generating 
functional of the Green functions of the (regularized) theory.
By $\langle\ ,\  \rangle$ we denote the scalar product in 
$L_2(\bbbr^4, d^4 x)\,$ so that
the second term contains the 0-loop two-point function.  Thus
$\Lll(\phi)\, $ generates the CAG, apart from the order zero contribution
given by the inverted free propagator. 
The functional $\Llol(\phi)= L^0(\phi)\,$ is the bare action including
counterterms, to be calculated from the renormalization conditions. 
On adding the 0-loop two-point function
and including 
the $\phi ^6$-insertion it takes the form (see (\ref{phi4}, \ref{phi41})) 
\eq
L'^{0}\,=\,
\int\biggl({Z\over 2} (\partial_{\mu}\phi)^2(x)\,+\,
{Z\over 2} (m^2+\de m^2) \phi^2(x)\,+\,
{g_0 \over 4~!} Z^2\,\phi^4(x)\,+\,{Z^3 \over \Lao ^2} 
{g_6\over 6~!}  \phi^6(x)\biggr)
\;d^4x\ .
\label{nawi}
\eqe
Here $Z\,,\ \de m^2\,$ and $g_0\,$ are formal power series in $\hbar\,$. 
The Wilson flow equation (FE)  
is is a differential equation for the functional
$L^{\La,\Lao}\,$, obtained from (\ref{funcin}) on differentiating 
w.r.t. $\La\,$~:
\eq
\partial_{\La}(\Lll + I^{\La,\Lao} ) \,=\,\frac{\hbar}{2}\,
\langle\frac{\delta}{\delta \phi},(\partial_{\La}{\hat C}^{\La,\Lao})
\frac{\delta}{\delta \phi}\rangle\Lll
\,-\,
\frac{1}{2}\, \langle \frac{\delta}{\delta
  \phi}\Lll,(\partial_{\La}
{\hat C}^{\La,\Lao})
\frac{\delta}{\delta \phi} \Lll\rangle \;\,.
\label{feq}
\eqe
Changing to momentum space and
expanding in a formal powers series w.r.t. $\hbar$ we write
(with slight abuse of no\-tation)
\eq
L^{\La,\Lao}(\vp)\,=\,\sum_{l=0}^{\infty} \hbar^l\,L^{\La,\Lao}_{l}(\vp)\,.
\label{3.3}
\eqe
From $L^{\La,\Lao}_{l}(\vp)$ we then obtain the CAG of loop order $l$
in momentum space  as 
\footnote{The  normalization of the ${\cLlln}$ 
is defined differently from earlier references.}
\eq
(2 \pi)^{4(n-1)} \de_{\vp(p_1)} \ldots \de_{\vp(p_n)}
L^{\La,\Lao}_l|_{\vp \equiv 0}
\,=\,
\de^{(4)} (p_1+\ldots+p_{n})\, {\cLlln}(p_1,\ldots,p_{n-1})\ ,
\label{cag}
\eqe
where we have written 
$\delta_{\vp(p)}=\delta/\delta\vp(p)$.
Note again that our definition of the $\cLlln$ is such that $\cLll_{0,2}$
vanishes. This is important for the set-up of the inductive scheme, 
through which perturbative renormalizability will be established.
The FE (\ref{feq}) rewritten in terms of the CAG (\ref{cag})
takes the following form
\eq
\pa_{\La} \pa^w \,\cLlln (p_1,\ldots, p_{n-1}) =
{1 \over 2} \int_k (\pa_{\La}C^{\La,\Lao}(k))\,\pa ^w 
\cLll_{l-1,n+2}(k,-k, p_1,\ldots, p_{n-1})\,-
\label{fequ}
\eqe
\[
-\!\!\!\!\!\!\!\!\!\!\!\sum_{l_1+l_2=l,\,w_1+w_2+w_3=w\atop
n_1+n_2=n} \!{1 \over 2} 
\Biggl[ \pa^{w_1} \cLll_{l_1,n_1+1}(p_1,\ldots,p_{n_1})\,
(\pa^{w_3}\pa_{\La}C^{\La,\Lao}(p'))\,\,
\pa^{w_2} \cLll_{l_2,n_2+1}(p_{n_1+1},\ldots,p_{n})\Biggr]_{ssym}\!\!\!,
\]
\[ \mbox{where }\quad
p'= -p_1 -\ldots -p_{n_1}\,= \,p_{n_1+1} +\ldots +p_{n}\ .
\] 
Here we have written (\ref{fequ})  directly in a form
where also momentum derivatives of the  CAG (\ref{cag})
are performed, and we used the shorthand notation
\eq
\pa^w:= \prod_{i=1}^{n-1}\prod_{\mu=0}^{3}
({\pa \over \pa p_{i,\mu}})^{w_{i,\mu}}\ \mbox{ with }\
w=(w_{1,0},\ldots,w_{n-1,3}),\  |w|=\sum w_{i,\mu} \in \bbbn_0\ .
\eqe 
The symbol $ssym$ 
(as defined in [KMR])
 means summation over those permutations of the momenta 
$p_1,\ldots, p_n$, which do not leave invariant the subsets
$\{p_1,\ldots, p_{n_1}\}$ and $\{p_{n_1+1},\ldots,p_{n}\}$.
Note that the CAG are symmetric  in their momentum arguments by definition.
A simple inductive proof of the renormalizability of $\phi_4^4$ theory 
has been exposed several times in the literature [KKS, KeKo1, Kop], 
and we will not repeat it in detail. 
The line of reasoning can be resumed as follows.
The induction hypotheses to be proven are~:\\
A) Boundedness
\eq
|\pa^w \cLlln(\vec{p})| \le\,
(\La+m)^{4-n-|w| }\,{\cal P}(log {\La + m \over m})\,
{\cal P}({|\vec{p}| \over \La+m})\ .
\label{propo1}
\eqe
B) Convergence
\eq
|\pa_{\Lao} \pa^w \cLlln(\vec{p})| \le\,
{1\over \Lao^3}{\cal P}(log {\Lao  \over m})\, (\La+m)^{6-n-|w| }\,
{\cal P}({|\vec{p}| \over \La+m})\ .
\label{propo2}
\eqe
Here and in the following the ${\cal P}\,$ denote 
(each time they appear possibly new)
polynomials with nonnegative coefficients. 
The coefficients 
depend on $l,n,|w|,m$,
but not on $\vec{p},\,\La,\,\Lao$. We used the shorthand
$\vec{p}=(p_1,\ldots,p_{n-1})$ and $|\vec{p}|=\sup
\{|p_1|,\ldots,|p_n|\}$.
The statement (\ref{propo2}) implies renormalizability~:
It proves the limits
\(\,\lim_{\La \to 0, \Lao \to \infty}\) \(  \cLlln(\vec{p})\,\) to exist
to all loop orders $l\,$.
But the statement (\ref{propo1}) has to be obtained first
to prove (\ref{propo2}). 
To prove (\ref{propo1}) we use an
inductive scheme that proceeds upwards in $2l+n$,
for given $2l+n$ upwards in $l$, and for given $(l,n)$ downwards in $|w|$, 
starting from some arbitrary $|w_{max}| \ge 3\,$. 
The important point to note is that the terms on the r.h.s. of the FE
 (\ref{fequ}) always are prior to the ones
 on the l.h.s. in the inductive order.
So the bound (\ref{propo1}) may be used as an induction hypothesis on
the r.h.s. Besides we also need a bound on the propagator and its
momentum derivatives~: It is easy to prove that
\eq
|\pa ^w \pa_{\La} C^{\La,\Lao}(p)| \le \La ^{-3-|w|} \,{\cal P}(|p|/\La)\,
e^{-{p^2+m^2 \over \La ^2}}\ .
\label{bopro}
\eqe 
Equipped with this bound and the induction scheme,
we may then integrate the FE, 
where terms with $n+|w| \ge
5\,$ are integrated down from $\Lao\,$ to $\La\,$, since for those terms we
have the boundary conditions following from 
(\ref{nawi})\footnote{Strictly speaking the boundary
condition for the $\cLlol_{l,6}$ has to be chosen more generally
first~: $\cLlol_{l,6} \le g_6\,
{1 \over \Lao ^2} {\cal P}(log {\Lao\over m})\, $. Once the bounds
on $\cLlol_{l,2}\,$ have been established
(which are independent of $\cLll_{l,6} \,$ at the same loop order), 
we specialize to $Z^3\,$
knowing that $Z^3\,$ is bounded by ${\cal P}(log{\Lao\over m})\,$.}
\eq 
\pa ^w \cLlol_{l,n} \equiv 0 \ \mbox{ if } \ n+|w|>4 \ \mbox{ and
  }  \ n\neq 6\,,
\ \mbox{ and } \ 
\cLlol_{l,6} \equiv  g_6\,{1 \over \Lao ^2} Z_l^3 \ .
\label{bo1}
\eqe 
The relevant terms (those with $n+|w|\le 4\,$) are integrated upwards 
from $0\,$ to $\La\,$. The boundary conditions for these terms are 
the renormalization conditions we impose and which fix the counterterms 
$Z, \ \de m^2,\ g_0\,$. We may fix for example
\eq 
\cLol_{l,2} (0)= 0\,, \
\pa_{p^2} \cLol_{l,2}(0)= 0\,, \
\cLol_{l,4}(0)= g \,\de_{l,0} \ .
\label{bo2}
\eqe 
To go away from the renormalization point (here chosen at zero
momentum) we may use the Schl{\"o}milch or integrated Taylor formula
which takes us back to the irrelevant situation.\\  
The bound (\ref{propo2}) holds for the $\bbbz_2$-symmetric theory
only, it is sharper than the one from [KKS,KeKo1,Kop]
(but generally assumed true in the literature).
Its proof is based on the same inductive scheme as 
that for (\ref{propo1}).
We start from the FE (\ref{fequ}) with $|w|$ momentum derivatives 
applied on it, integrate over $\La$ and derive w.r.t. $\Lao$.
For the terms on the r.h.s., on which this derivative does not apply,
we can use the bound (\ref{propo1}). For the terms derived w.r.t.
$\Lao$ we can use (\ref{propo2}), applying our induction
 scheme. The best bound we can arrive at is essentially    
saturated by the boundary terms, we find. \\
We first regard the case of the irrelevant terms with $n+|w| \ge 5$,
and here we start looking at the case $n+|w| \ge 6$. We have the
equation (in shorthand notation)
\eq
-\pa_{\Lao} \pa ^w \cLll_{l,n}=
\pa ^w (\mbox{r.h.s. of the FE})|_{\La =\Lao}+\int_{\La}^{\Lao} \!\!\!\!\!\!
d\La'
\pa_{\Lao} \pa ^w (\mbox{r.h.s. of the FE})(\La')\ .
\label{felao}
\eqe
Now using (\ref{propo1}, \ref{propo2}) to bound the r.h.s. of
(\ref{felao}) we verify  (\ref{propo2})  on performing the integral.
In the case $n+|w| = 5$ it suffices to regard 0 external momentum
(referring again to the Schl{\"o}milch formula for deviations from 0,
which takes us back to $n+|w| = 6\,$). 
In both cases $n=4, |w| = 1$ and $n=2, |w| = 3$ the first term of the
r.h.s.
of (\ref{felao}) is 0 due to our boundary conditions,
whereas the second vanishes due to  euclidean invariance.\\
Terms with $n+|w| \le 4$ have to be analysed at the renormalization point
indicated in (\ref{bo2}), and they are integrated from 0 to $\La$~:
\eq
\pa_{\Lao} \pa ^w \cLll_{l,n}=
\int_{0}^{\La} d\La'\,
\pa_{\Lao} \pa ^w (\mbox{ r.h.s. of the FE })(\La')\ .
\label{felao1}
\eqe 
Only the second term from (\ref{felao}) appears on the r.h.s. 
because the renormalization
conditions are $\Lao$-independent. In this term we may factorize
$\Lao ^{-3} {\cal P}(log \Lao)\,$ and verify (\ref{propo2}) by induction.
As a result of these considerations we obviously obtain 
the same bounds for the theory with an insertion of $\phi^6\,$ as 
for the theory with $g_6=0\,$. \\

To study the theory in the critical domain, 
and particularly the role of the irrelevant
insertions in this domain, we have to analyse the correlation
functions in the limit of large correlation length, i.e. in the
language of field theory, (the approach to) the massless theory.    
Thus we regard the propagator
\eq 
C^{\La,\Lao}(p)\,=\, \frac{1}{p^2}(e^{-\frac{p^2}{\Lao^2}}\,-\,e^{-\frac{p^2}{\La^2}})
\,,\quad (\La,p)\neq (0,0)\,,
\eqe
which is singular for $\,\sup(\La ^2,p^2 )\to 0$.
However it has a finite limit for $\,p^2 \to 0$, if $\La$
stays bounded from below so that (\ref{bopro})
stays valid for $\La \ge m\,$. Only the case $\La <m\,$ has to be
reconsidered.
 
For $\La \to 0\,$ in fact
the CAG may become singular at certain
exceptional momentum configurations, i.e.
where subsums of external momenta vanish.
But first, for the massless theory to exist at all, certain restrictions on the
renormalization conditions have to be observed. More specifically
the renormalization points have to be chosen as follows~:
\eq
\ccL^{0,\Lao}_{l,2}(0)= 0\,,\ \,
(\partial_{p_{\mu}}\partial_{p_{\nu}}
\ccL^{0,\Lao}_{l,2})(p=k)|_{\de_{\mu,\nu}}
=0\,,\ \,
\ccL^{0,\Lao}_{l,4}(k_1,k_2,k_3)=g \,\de_{l,0}\,\;.
\label{rencon}
\eqe
This means the mass renormalization has to be performed at 0 momentum,
whereas the wave function renormalization and coupling constant
renormalization have to be performed at nonexceptional external
momenta, i.e. $k^2 =\mu ^2 \neq 0$ and no subsum in $k_1,\,k_2,\,k_3,\,k_4$
vanishes. Since we have defined the $\cal L$ to be 
symmetric
functions of their arguments it is natural to make a  
symmetric\footnote{But it is not
  necessary because the solutions of the FE come out symmetric
by construction.} choice, e.g.  
$\vec{k_i} \cdot \vec{k_j}= {\mu ^2 \over 4} (4 \de_{ij} -1)$ for
a fixed nonvanishing momentum scale $\mu$.
In (\ref{rencon})
$\,(\partial_{p_{\mu}}\partial_{p_{\nu}}
\ccL^{0,\Lao}_{l,2})(p=k)|_{\de_{\mu,\nu}}$ 
refers to the decomposition of the O(4) invariant tensor  
\[
\partial_{p_{\mu}}\partial_{p_{\nu}}\cLol_{l,2}(p)|_{p=k}\,=\,
A(\mu^2)\de_{\mu,\nu}\,+\, B(\mu^2)k_{\mu}k_{\nu}\,,
\]
and we have defined 
\eq
\partial_{p_{\mu}}\partial_{p_{\nu}}\cLol_{l,2}(p=k)|_{\de_{\mu,\nu}}\,=\,
A(\mu^2)
\eqe
so that the renormalization condition  implies
$A(\mu^2)\,=\,0\,$. Note that $B(\mu^2)$ is irrelevant and need not be
fixed by a renormalization condition. Obviously renormalization of the
massless theory introduces a new mass scale which is generally 
called $\mu$.
The problem of exceptional momentum configurations can be studied
in full generality and rigour with flow equations [KeKo2]~: 
It is possible to
define an IR index $\ga$ with $2\ga \in  \bbbn$, which measures the
exceptionality of the momentum configuration $P =(p_1,\ldots,p_n)\,$. 
Using the  shorthand notations
$\,\ccL =\ccL^{0,\infty}$ and $\ccL^{\La}=\ccL^{\La,\infty}\;$,
we may phrase as follows the results from 
the  renormalization proof [KeKo2] for the
massless symmetric $\phit$ theory~:

a) The n-point CAG with for $n>2$ are smooth functions of the external
momenta in the (open) subspace of 
(arbitrarily) bounded
nonexceptional momentum configurations.  
We have 
\eq
\partial^w \ccL_{l,n}(p_1,\ldots,p_{n-1})\,=\,
\lim_{\La\to 0} \partial^w \ccL_{l,n}^{\La}(p_1,\ldots,p_{n-1})\ .
\eqe
b) Generally one has
\eq
|\partial^w \cLl_{l,n}(p_1,\ldots,p_{n-1})|\,\leq\,
\mu^{4-n-|w|}\, \bigl({\mu \over \La}\bigr)^{2\ga(P)+|w|}\, 
{\cal P}(log {\mu \over \La})\;,
\quad 0<\La\leq \mu\ .
\label{exc}
\eqe
For the two point function at $\La =0$ one can also show that it vanishes
as $\,O \bigl(p^2 \,{\cal P}(\log {\mu^2\over p^2})\bigr) \,$ 
near $0\,$ momentum. 

Since $\La\,$ acts as an infrared regulator the bounds 
(\ref{propo1}, \ref{propo2}) still hold for 
$\La>\mu\,$, on replacing $m\,$ by $\mu\,$.  
For $\La < \mu\,$ these bounds also hold for nonexceptional momentum
configurations. For exceptional configurations they have to be
multiplied by the power of $\mu/\La\,$ appearing in (\ref{exc}). 
We do not enter into details of the infrared problem, since 
the bounds in the region
$\La <\mu\,$ are independent of the $\phi^6\,$-insertion
and therefore the proof from [KeKo2] may be taken over unaltered.
As regards the term 
$\Lao ^{-3} {\cal P}(log {\Lao\over \mu})\,$ appearing in (\ref{propo2}),
it does not interfer with the exceptional momentum problem and
can be factored out in the inductive proof as it was done  before 
in the massive case.

We now want to establish bounds for the 
difference between the theories with and without
$\phi ^6\,$-insertion. The CAG of this theory are to be called 
$\De \cLll_{l,n}\,$. This means we define (in obvious notation)
\eq
\De \cLll_{l,n}(p_1,\ldots,p_{n-1})=
\cLll_{l,n}(g_6~;p_1,\ldots,p_{n-1})-
\cLll_{l,n}(0~;p_1,\ldots,p_{n-1})\ .
\label{del}
\eqe 
Here it is understood that the $\cLll_{l,n}(g_6)\,$ and 
the $\cLll_{l,n}(0)\,$ obey the {\it same} renormalization conditions,
which means that  
all the relevant $\De \cLll_{l,n}\,$ are imposed to vanish 
at the renormalization point.
We may obtain  the flow equations for  the $\De \cLll_{l,n}\,$
by taking the difference between those for 
$\cLll_{l,n}(g_6)\,$ and $\cLll_{l,n}(0)\,$. We only give it in 
shortened form without momentum arguments, 
the explicit form following directly
from (\ref{fequ}). We get  
\eq
\pa_{\La} \pa^w \,\De \cLlln =
{1 \over 2} \int_k (\pa_{\La}C^{\La,\Lao}(k))\,\pa ^w 
\De \cLll_{l-1,n+2}(k,-k,\ldots)\,-
\label{fequ1}
\eqe
\[
-\!\!\!\!\!\!\!\!\!\!\!\sum_{l_1+l_2=l,\,w_1+w_2+w_3=w\atop
n_1+n_2=n} \!{1 \over 2} 
\Biggl[ \pa^{w_1}\De  \cLll_{l_1,n_1+1}\,
(\pa^{w_3}\pa_{\La}C^{\La,\Lao})\,\,
\pa^{w_2}\bigl(
 \cLll_{l_2,n_2+1}(g_6)\,+\,\cLll_{l_2,n_2+1}(0)\,\bigr)
\Biggr]_{ssym}\ .
\]
With this system of equations we can inductively prove the following
bounds for the massless theory. For nonexceptional momentum
configurations one finds 
\eq
|\partial^w \De \cLll_{l,n}(\vec{p})|\,\leq \,
\left \{\!\! \begin{array}{r@{\,,}l} 
{{\cal P}(\log {\Lao \over \mu} ) \over \Lao ^{2}}\,\,\mu^{6-n-|w|}\,\, 
{\cal P}({|\vec{p}| \over \mu})  
& \ \mbox{ for }
\quad 0 \leq \La \leq \mu     \\  
{{\cal P}(\log {\Lao \over \mu} ) \over \Lao ^{2}}\,\, 
\La^{6-n-|w|}\,\,
{\cal P}(\frac{|\vec{p}|}{\La})
 & \ \mbox{ for } \ \,\, \mu\leq \La \leq \Lao\;\\
\end{array}  \right \}\ ,
\label{bound1}
\eqe
whereas for general momentum configurations one obtains
\eq
|\partial^w \De \cLll_{l,n}(\vec{p})|\,\leq \,
\left \{\!\! \begin{array}{r@{\,,}l} 
{{\cal P}(\log {\Lao \over \mu} ) \over \Lao ^{2}}\,\, 
{\cal P}(\log \mu/\La)\,\mu^{6-n-|w|}\,\, 
({\mu \over \La})^{2\ga+|w|}\,
{\cal P}({|\vec{p}| \over \mu})  
& \ \mbox{ for }
\quad 0 \leq \La \leq \mu     \\  
{{\cal P}(\log {\Lao \over \mu} ) \over \Lao ^{2}}\,\, \La^{6-n-|w|}\,\,
{\cal P}(\frac{|\vec{p}|}{\La})
 & \ \mbox{ for } \ \,\, \mu\leq \La \leq \Lao\;\\
\end{array}  \right \}\ .
\label{bound2}
\eqe
We do not give a proof of these bounds, since they are 
obtained using the same inductive scheme as before,
applying also the bounds for 
$\cLll_{l,n}(g_6)\,$ and $\cLll_{l,n}(0)\,$ obtained previously.
The improvement factor ${\cal P}(\log {\Lao \over \mu} ) / \Lao ^{2}\,$
is respected in particular by the new boundary conditions: 
All renormalization
conditions vanish, and the only nonvanishing boundary term, i.e.
the term $\sim g_6 \,Z^3/ \Lao ^{2}\,$ for the six point function 
satisfies (\ref{bound1}, \ref{bound2}). 
Still we would like to point out that rigorous bounds as (\ref{propo2}, 
\ref{bound1}, \ref{bound2}) are hard (if not impossible) to
obtain by other methods. We will use them in the next section
to obtain equivalent bounds on the corrections to scaling
due to irrelevant terms. 

\section {Renormalization Group Equations and Critical Behaviour}

We will use the previous results to analyse
the modification of critical behaviour
by irrelevant terms without composite operator formalism.
The advantages of this procedure have been mentioned before.
In this last section we will change to the standard notation in the sense
that now ${\cal L}^{0,\Lao}_2\,$ denotes the two point function 
{\it including} the 0 loop contribution. 
Our CAG $n\,$-point functions ${\cal L}^{0,\Lao}_n\,$ are defined in 
terms of the field variable
$\phi\,$, which is the renormalized field in standard language.
Relating them to the bare functions expressed in terms of the bare
field $\phi_B$ which is related to $\phi$ through the relation
\eq
\phi_B =  Z^{1/2} \phi
\eqe
we obtain 
\eq
{\cal  L}^b_n(p_i,g_0,g_6,\Lao)=Z^{n/2}(g,g_6,{\Lao \over \mu} )\, 
{\cal L}^{0,\Lao}_n(p_i,g,g_6,\mu)\ .
\label{mult}
\eqe  
The sign in the exponent of $Z$ is
related to the fact that the functions ${\cal L}^{0,\Lao}$ are
the connected {\it free} propagator amputated functions. 
This  sign changes if we
use the full propagator amputated functions instead, which is 
of course possible, but less natural in the FE framework.
Taking a derivative of (\ref{mult}) w.r.t. $\ln \mu$ at fixed
bare parameters we obtain  the (standard) renormalization
group equation for the renormalized theory
\eq
\left [\frac{\partial}{\partial \ln \mu}+\beta(g,g_6;{\mu \over \Lao})
\,\frac{\partial}
{\partial g}+\frac{1}{2}n\,\gamma(g,g_6;{\mu \over \Lao}) \right]
{\cal L}^{0,\Lao}_n(p_i,g,g_6,\mu)= 0\ .
\label{rengr}
\eqe
We have introduced the  $\beta\,$ and $\ga\,$ functions for the 
renormalized theory
\eq
\beta(g,g_6;{\mu \over \Lao})=\frac{\partial\, g}{\partial\, \ln
  \mu}|_{g_0,g_6,\Lao}\,,\quad 
\gamma(g,g_6;{\mu \over \Lao})
=\frac{\partial\,\ln Z}{\partial \, \ln \mu}|_{g_0,g_6,\Lao}\,.
\eqe
Since we want to use this equation for large but nevertheless finite
$\Lao$ the functions $\beta(g,g_6;{\mu \over \Lao})$ 
and $\gamma(g,g_6;{\mu \over \Lao})$ depend also
on $\Lao\,$. 
Due to (\ref{propo2}) 
the $\Lao$-dependent terms are bounded by
$O(({\mu \over \Lao})^{-2} {\cal P}(\log {\Lao \over \mu}))\,$, since  
$\beta(g,g_6;{\mu \over \Lao})$ and $\gamma(g,g_6;{\mu \over \Lao})$ 
may be expressed in terms
of  ${\cal L}^{0,\Lao}$ using (\ref{rengr}) 
for fixed values of $n\,$: By dimensional analysis we transform
the derivative w.r.t. $\mu\,$ into a derivative w.r.t. $p\,$ and
$\Lao\,$ and obtain
from the equations for $n=4\,$ and for $n=2\,$~:
\[
\beta(g,g_6;{\mu \over \Lao})\,=
\]
\eq
=\,\sum_{i=1}^3 p_{i,\nu} 
{\pa \over \pa p_{i,\nu}} {\cal L}^{0,\Lao}_4(g,g_6)|_{r.p.}
\,-\,4 g\, p^2{\pa ^2 \over \pa (p^2)^2}{\cal L}^{0,\Lao}_2(g,g_6)|_{r.p.}
+O\bigl(({\mu \over \Lao})^{2} {\cal P}(log {\Lao\over \mu})\bigr)\, ,
\label{beta}
\eqe 
\[
\ga(g,g_6;{\mu \over \Lao})=
2\, p^2{\pa ^2 \over \pa (p^2)^2}{\cal L}^{0,\Lao}_2|_{r.p.}
+O\bigl(({\mu \over \Lao})^{2} {\cal P}(\log {\Lao\over \mu})\bigr)\ . 
\] 
The functions are to be taken at the renormalization points 
(see (\ref{rencon})). The contributions $\sim O(\Lao ^{-2} 
{\cal P}\log \Lao)\, $
arise when transforming  the  $\mu\,$-derivative into one on $\Lao\,$
on using the bound  
(\ref{propo2}).
So to be precise we rewrite (\ref{rengr}) as
\eq
\left [\frac{\partial}{\partial \ln \mu}+\beta(g,g_6)\,\frac{\partial}
{\partial g}+\frac{1}{2}n\,\gamma(g,g_6) \right]
{\cal L}^{0,\Lao}_n(p_i,g,g_6,\mu)= 
O\bigl(({\mu \over \Lao})^{2} {\cal P}(\log {\Lao\over \mu})\bigr)\; ,
\label{rengr1}
\eqe
where the whole dependence on $\Lao\,$ has been regrouped on the r.h.s.
(with the definitions 
$\beta(g,g_6)= \beta(g,g_6;0)\,, \ \ga(g,g_6)= \ga(g,g_6;0)\,$).
When setting $g_6\,=0\,$ we obtain the corresponding equation 
with functions  $\beta(g,0;{\mu \over \Lao})\,$ and 
$\gamma(g,0;{\mu \over \Lao})\,$ obeying the equations
analogous to  (\ref{beta})  for $g_6\,=0\,$. From this it follows
on using (\ref{bound1}) that 
\eq
\De \be(g,g_6;{\mu \over \Lao})~:=
\be(g,g_6;{\mu \over \Lao})-\be(g,0;{\mu \over \Lao})=
O\bigl(({\mu \over \Lao})^{2} {\cal P}(\log {\Lao\over \mu})\bigr)\ ,
\label{borne}
\eqe
and similarly for $\ga\,$. This bound
can of course be verified in lowest orders
by direct calculation of the respective
$\be\,$-functions.  When the $\phi^6\,$-term is added, 
the two diagrams given in Fig.1 contribute to 
the relation between $g\,$ and $g_0\,$ and thus to 
$\,\be(g,g_6;{\mu \over  \Lao})\,$ up to two loops.
\footnote{We did not include those diagrams which are exactly cancelled 
by  diagrams carrying an insertion of a counterterm.} Since the
second diagram is $\mu\,$-independent, only the first contributes 
to the $\be\,$-function. The value of the diagram 
is
\[
g\,g_6\;{2\over 16 \pi ^4}\,\ln {4\over 3} \,+\,O\bigl(({\mu
  \over \Lao})^{2} \log ({\Lao\over \mu})\bigr)\; , 
\]
so that after derivation w.r.t. $\ln \mu\,$ its contribution is of the
order given in (\ref{borne}).\\[.5cm]  

\begin{picture}(200,100)(-100,0)

\put(0,60){\line(1,0){80}}
\put(60,60){\line(1,1){20}}
\put(60,60){\line(1,-1){20}}
\put(40,60){\circle{40}}
\put(20,60){\circle*{4}}
\put(60,60){\circle*{4}}
\put(15,43){$g$}
\put(60,43){$g_6$}

\put(120,60){\line(1,0){40}}
\put(140,60){\line(-1,-1){20}}
\put(140,60){\line(1,-1){20}}
\put(140,80){\circle{40}}
\put(140,60){\circle*{4}}
\put(135,43){$g_6$}

\end{picture}

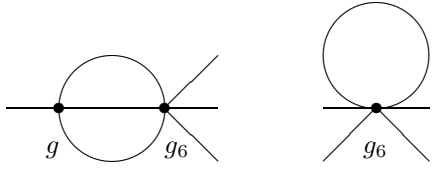
{ \sl Figure 1: Contributions $\sim g_6\,$ to the relation between
$g\,$ and $g_0\,$ up to two loops.}\\[.5cm]

We refer to the textbooks [ZJ, IZ] for the method of solution of 
(\ref{rengr}), which permits to compare 
${\cal L}^{0,\Lao}_n(p_i,g,g_6,\mu)\,$ to
${\cal L}^{0,\Lao}_n(\frac{p_i}{s},g(s),g_6,\mu)\,$, 
the critical region   
corresponding to $s \rightarrow \infty\,$. 
Here the running coupling at scale $\mu/s\,$ is defined through
\eq
\frac {dg(s)}{d \, ln \, s}=-\beta(g(s))\,; \quad g(1)=g\ .
\eqe
From (\ref{rengr}), together with dimensional analysis,    
one obtains 
\eq
{\cal L}^{0,\Lao}_n(p_i,g,g_6,\mu)=s^{-4+n} \,
e^{\frac{1}{2}n\int^{g(s)}_{g}\!\!\frac{\gamma (g',g_6;\mu/\Lao) }
{\beta  (g',g_6;\mu/\Lao)}dg'}\,\,
{\cal L}^{0,s\Lao}_n(s  p_i,g(s),g_6,\mu) 
\eqe
\noindent
or on replacing $p_i$ by $p_i/s$:
\eq
{\cal L}^{0,\Lao}_n(\frac{p_i}{s},g,g_6,\mu)=s^{-4+n}\,
e^{\frac{1}{2}n\int^{g(s)}_{g}
\frac{\gamma  }{\beta } dg'}
{\cal L}^{0,s\Lao}_n(p_i,g(s),g_6,\mu)\ .  
\label{rgg}
\eqe

For $s>\!>1\,$ the coupling will approach its fixed point value
$g^*\,$ for which by definition $\beta(g^*) =0\,$. In the
perturbative region we have $g^*=0\,$ in $d=4\,$ , whereas in $d<4\,$ 
one finds 
$g^*=O(\vep)\,$ with $\vep=4-d\,$. If $g\,$ is in the vicinity of the fixed
point the integral
$\int^{g(s)}_{g}\! \!\frac{\gamma (g') }{\beta (g')}\,dg'\,$ is approximated
by its value at $g^*\,$ 
\eq
-\int^{g(s)}_{g}\!     
\frac{\gamma
  (g')}
{\beta (g') }\,dg'\,=\int^{\ln
  \,s}_{0} \!\! \! 
\gamma(g(s'))  \,d\ln s'\
\sim\,\gamma(g^*)\,\ln s\ .
\eqe
The neglected terms give subdominant contributions for $s \to \infty\,$,
they are analysed in [BGZ].
From this we then find for the dominating behaviour
\eq
{\cal L}^{0,\Lao}_n(\frac{p_i}{s},g,g_6,\mu)\sim\, 
s^{-4+n(1- {\ga(g^*)\over     2})}\,
{\cal L}^{0,\infty}_n(p_i,g^*,g_6,\mu)\, ,
\eqe
which shows that the fixed point value $\gamma(g^*)$ is to be
identified with the critical exponent $\eta$.

The renormalization group equation for the difference functions 
(\ref{del}) can be obtained from
(\ref{rengr}). We write it in the form
\eq 
\left [\frac{\partial}{\partial \ln \mu}+
\beta(g,0;{\mu \over \Lao})\,\frac{\partial}
{\partial g}+\frac{n}{2}\,\gamma(g,0;{\mu \over \Lao}) \right]
\De {\cal L}^{0,\Lao}_n(p_i,g,g_6,\mu)\,=
\label{derengr}
\eqe
\[
= \,
-\left [\De\beta(g,g_6;{\mu \over \Lao}) 
\frac{\partial}{\partial g} + {n \over 2} 
\De\ga(g,g_6;{\mu \over \Lao})\right]\,
{\cal L}^{0,\Lao}_n(p_i,g,g_6,\mu)\ .
\]
For the inhomogeneous equation we make the ansatz 
\eq
\De {\cal L}^{0,\Lao}_n(p_i,g,g_6,\mu)= U^{0,\Lao}_n(p_i,g,g_6,\mu)
\,{\cal L}^{0,\Lao}_n(p_i,g,0,\mu)\ .
\label{dec}
\eqe
From this we obtain the following differential equation for $U^{0,\Lao}\,$ 
\eq
(\frac{\partial}{\partial \ln \mu}+\beta(g,0;{\mu \over \Lao})\,
\frac{\partial}{\partial g})\,
U^{0,\Lao}\,=
\eqe
\[
=\,-[\De\beta(g,g_6;{\mu \over \Lao}) \frac{\partial}{\partial g}\ln{\cal
  L}_n^{0,\Lao}(p_i,g,0,\mu) 
 + {n \over 2} \De\ga(g,g_6;{\mu \over \Lao})]\,+\,
O\bigl(({\mu \over \Lao})^{-4} \,{\cal P}(\log {\Lao\over \mu}) \bigr)\ .
\]
In the following we will negelect the last term which gives even
smaller corrections, for the first two terms on the r.h.s. of 
this equation we write 
$V_n(p_i,\mu,g;g_6,\Lao)\,$.
Its solution is then obtained as a sum of the general
soultion of the corresponding homogeneous equation -which in turn is
obtained as previously for the case $\ga=0\,$- 
plus a special solution of the inhomogeneous equation, which can be
written as the integral over $V_n(p_i,\mu,g;g_6,\Lao)\,$. 
As a final result we obtain the following renormalization group
relation for 
$U^{0,\Lao}_n(p_i,g,g_6,\mu)$~:
\eq
U^{0,\Lao}_n(p_i,g,g_6,\mu)=
U^{0,\Lao}_n(p_i,g(s),g_6,\mu/s)+\,
\int_{-ln\,s}^0 
V_n(p_i, \mu e ^{t},\,  g(e ^{-t});g_6,\Lao)\, dt\ .
\eqe
By dimensional analysis we obtain
\eq
U^{0,\Lao}_n(p_i,g,g_6,\mu)=
U^{0,s\Lao}_n(s\,p_i,g(s),g_6,\mu)+\,
\int_{-ln\,s}^0 
V_n(s p_i,s \mu e ^{t}, g( e ^{-t});\,g_6,s \Lao)\, dt\ ,
\eqe
since the canonical dimension of $U_n\,$ is zero. Multiplying by
$\,{\cal L}^{0,\Lao}_n(p_i,g,0,\mu)\,$,  using (\ref{rgg}) and passing
to momenta $p_i/s\,$ we thus obtain
\eq
\De {\cal L}^{0,\Lao}_n({p_i \over s},g,g_6,\mu)= 
s^{-4+n}e^{\frac{1}{2}n\int^{g(s)}_{g}
\frac{\gamma (g',0;\mu/\Lao) }
{\beta  (g',0;\mu/\Lao)} dg'}
 \biggl[\De {\cal L}^{0,s\Lao}_n(p_i,g(s),g_6,\mu)\,+\,
\eqe
\[ +
{\cal L}^{0,s\Lao}_n(p_i,g(s),0,\mu)\,
\int_{-ln\,s}^0\!\!\! 
V_n(p_i,s\mu e^{t},g( e^{-t});\,g_6,s\Lao)\, dt \biggr]\
.
\]
The second  term can be bounded using  
(\ref{borne}) (together with 
(\ref{propo1}))\footnote{It is useful to cut the integration interval
into subintervals of length $\,\ln 2\,$ and sum  over 
the bounds for the integrand in the subintervals to avoid a factor of
$\ln s\,$ in the bound for this term.},  
to the first term  we can apply (\ref{bound1}) to obtain the
following bound on $\,\De {\cal L}^{0,\Lao}_n({p_i \over
  s},g,g_6,\mu)\,$: 
\eq
|\De {\cal L}^{0,\Lao}_n({p_i \over s},g,g_6,\mu)|
\le 
\eqe
\[
s^{-4+n}e^{\frac{1}{2}n\int^{g(s)}_{g}
\frac{\gamma }
{\beta  } dg'}\biggl[
O\bigl(({\mu \over s\Lao})^{2} {\cal P}(\log {s\Lao\over \mu})\bigr)\,+\,
|{\cal L}^{0,s\Lao}_n(p_i,g(s),0,\mu)|\;
O\bigl(({\mu \over \Lao})^{2}\,{\cal P}(\log {\Lao\over \mu})\bigr)
\biggr]\ .
\]
This bound is dominated for $s\,$ large by the second term so that
we obtain 
\eq
|\De {\cal L}^{0,\Lao}_n({p_i \over s},g,g_6,\mu)|
\le 
s^{-4+n}e^{\frac{1}{2}n\int^{g(s)}_{g}
\frac{\gamma  }
{\beta  } dg'}\;
|{\cal L}^{0,s\Lao}_n(p_i,g(s),0,\mu)|\;
O\bigl( ({\mu \over \Lao})^{2}\,{\cal P}(log {\Lao\over \mu})\bigr)\ .
\eqe
The analysis of the prefactor is the same as 
for ${\cal L}^{0,\Lao}_n({p_i \over s},g,0,\mu)\,$.
Therefore, close to the critical region, the corrections 
of the long distance behaviour due to the irrelevant term
are of the relative order $\,O(({\mu \over \Lao})^{2})\,$
 up to logarithms. 
For this term to be negligeable we need of course $\mu <\!<\Lao\,$,
that is to say, the renormalized parameters are close to the critical
ones, which is a natural parametrization in the critical region.  
We emphasize  that the corrections to
scaling stem from the analysis of the terms vanishing for $\Lao \to
\infty\,$, which are often neglected  altogether in the literature.
In the composite operator analysis one finds instead
corrections $\sim s^{-2} {\cal P}(\log s)\,$, which would be smaller
for $s\,> {\Lao \over \mu} \,$. 
However the terms $\sim ({\mu \over \Lao})^{2}\,$ are always present,
though often neglected, so that the corresponding results only hold
up to $s\sim {\Lao \over \mu} \,$. For larger $s\,$ one has to readapt the
renormalization conditions at $\mu' <\!<\mu \,$.
In terms of the bare theory the
readaptation consists in adding new counterterms $\sim \phi ^2\,$
and $\sim \phi ^4\,$. This is well known from the treatment in the
composite operator formalism, where such terms are introduced
due to operator mixing. 

In conclusion we thus realize that in our approach the  corrections to
scaling due to irrelevant terms are suppressed by $\,O\bigl(({\mu \over
  \Lao})^{2}\bigr)\,$ to any order in the number of insertions. 
These irrelevant terms are introduced directly in the bare
action, keeping the renormalization conditions fixed. In composite 
operator theory, which is completely bypassed here, 
the coefficient of the $\phi^6$-term in the bare action 
is not suppressed by $\,({1 \over
  \Lao})^{2}\,$, correspondingly one does not obtain such a
suppression in the corrections to scaling. Instead, on subtracting
insertions of lower dimension, to be calculated from the relations for
operator mixing, 
 one obtains a
suppression factor $\sim s^{-2}\,$,
which is larger for $\,s< {\Lao \over   \mu}\,$,
 becomes of similar size
for $s \sim {\Lao \over  \mu}\,$ and unreliable beyond.\\[3cm] 
\noindent 
{\bf References:}
\begin{itemize}
\item[[Amit]$\!\!\!$] D.J. Amit:  Field Theory, the Renormalization Group and
Critical Phenomena, $2^{nd}$ ed. World Scientific, Singapur (1984).
\item[[BGZ]$\!\!\!$] E. Br{\'e}zin, J.C. Le Guillou and  J. Zinn-Justin,
 in:  Phase
  Transitions and Critical Phenomena, Vol. VI, Ch.3, C. Domb et M. S. Green,
  eds. Academic Press, N.Y. (1976). 
\item[[IZ]$\!\!\!$] C. Itzykson and J.B. Zuber: Quantum Field Theory,
Mc Graw Hill, N.Y. (1980). 
\item[[KeKo1]$\!\!\!$] G. Keller and Ch. Kopper. 
Perturbative Renormalization of
  Composite Operators via Flow Equations I. 
Commun.Math.Phys.{\bf 148}, (1992) 445-467.
\item[[KeKo2]$\!\!\!$] G. Keller and Ch. Kopper, 
Perturbative Renormalization
  of Massless $\phi^4_4$ with Flow Equations. Commun.Math.Phys.{\bf
    161}, (1994) 515-532. 
\item[[KKS]$\!\!\!$] G. Keller, Ch. Kopper, M. Salmhofer,
Perturbative Renormalization and Effective Lagrangians in $\phi_4^4\,$.
Helv.Phys.Ac\-ta {\bf 65}, (1991) 33-52.
\item[[KMR]$\!\!\!$] Ch. Kopper,  V.F.M. M{\"u}ller, Th. Reisz,
Temperature Independent Renormalization of Finite Temperature Field
Theory. Preprint hep-th/0003254. 
\item[[KoSm]$\!\!\!$] Ch. Kopper and V.A. Smirnov, Analytic
  Regularization and Minimal Subtraction of $\phi_4^4\,$ with Flow
  Equations. Z.Phys.{\bf C59}, (1993) 641-645.
\item[[Kop]$\!\!\!$] Ch. Kopper: Renormierungstheorie mit Flussgleichungen,
  Shaker Verlag, Aachen (1998). 
\item[[Pol]$\!\!\!$] J. Polchinski, Renormalization and Effective Lagrangians. 
Nucl.Phys.{\bf B231}, (1984) 269.
\item[[Weg]$\!\!\!$] F. Wegner, Corrections to Scaling Laws, 
Physical Review {\bf B5}, (1972) 4529-4536, see also
F. Wegner, in:  Phase
  Transitions and Critical Phenomena, Vol. VI, Ch.2, C. Domb et M. S. Green,
  eds. Academic Press, N.Y. (1976).
\item[[WiKo]$\!\!\!$] K. Wilson et J. B. Kogut, 
The Renormalization Group and the $\vep\,$-Expansion.
Physics Reports, 12c, (1974) 77.
\item[[Zim]$\!\!\!$] W. Zimmermann, Composite Operators in the
  Perturbation Theory of Renormalizable Interactions,
Ann.Phys. {\bf 77}, (1973) 536-569.
\item[[ZJ]$\!\!\!$] J. Zinn-Justin:  Quantum Field Theory and Critical
Phenomena, Clarendon Press, Ox\-ford (1989).
\end{itemize}

\end{document}